\documentclass[referee]{aa} 
\usepackage{graphicx}
\begin{document}
   \title{Carbon star survey in the Local Group. VII. NGC 3109 a
 galaxy without a stellar halo
}


   \author{S. Demers
          \inst{1},
\fnmsep\thanks{Visiting Observers, 
   Canada-France-Hawaii Telescope, operated by the National Research 
Council of Canada, the Centre National de la Recherche Scientifique de 
France, and the University of Hawaii}
          P. Battinelli \inst{2}
 \and
B. Letarte\inst{1}\fnmsep\thanks{Current address:
                Kapteyn Astronomical Institute, Posbus 800, Groningen, 9700 AV,
        Netherlands}
          }

   \offprints{S. Demers}

   \institute{D\'epartement de Physique, Universit\'e de Montr\'eal,
                C.P.6128, Succursale Centre-Ville, Montr\'eal,
                Qu\'ebec H3C 3J7, Canada\\
                \email {demers@astro.umontreal.ca }
                \email {bruno@astro.rug.nl}
         \and
INAF, Osservatorio Astronomico di Roma
              Viale del Parco Mellini 84, I-00136 Roma, Italia\\
              \email {battinel@oarhp1.rm.astro.it }
}
   \date{Received; accepted}

   \abstract{We present a CFH12K wide field survey of the carbon star 
population in and around NGC 3109. Carbon stars, the brightest members
of the intermediate-age population, were found nearly exclusively in 
and near the disk of NGC 3109, ruling out the existence of an 
extensive intermediate-age halo like the one found in NGC 6822. 
Over 400 carbon stars identified have $\langle M_I\rangle = -4.71$,
confirming the nearly universality of mean magnitude of C star populations
in Local Group galaxies.
Star counts over the field reveal that NGC 3109 is a truncated 
disk shaped galaxy without an extensive stellar halo. The minor axis
star counts reach the foreground density between 4$'$ and 5$'$, a distance
that can be explained by an inclined disk rather than a spheroidal halo.
We calculate a global C/M ratio of 1.75 $\pm$ 0.20, a  value
expected for such a metal poor galaxy.
 }
\titlerunning{NGC 3109}

   \maketitle
%

\section{Introduction}

Halos of galaxies may result from gravitational collapse (Eggen et al.
 1962), early in their life, or from the accretion of small fragments
(Searle \& Zinn 1978). The halo building by accretion should be a continuous
process leading to halo populations of different ages and metallicities.
Recently, dwarf galaxies have been found to have halos of old stars (Lee 
1993; Minniti \& Zijlstra 1997; Minniti et al.  1999); or
stars older than 1.5 Gyr ( Aparicio \& Tikhonov 2000; Aparicio et al. 2000). 
In this regard, our
recent CFH12K observations of NGC 6822
(Letarte et al. 2002)  represent a major breakthrough. 
We detected numerous carbon stars (C stars)
in the halo of this Magellanic-type galaxy implying the presence of
an intermediate-age population extending at least three times the optical
radius. Albert et al.  (2000) have also identified C stars at large radii
in IC 1613. Do all Magellanic type galaxies have an intermediate-age
population in their halo?    Is there something special about NGC 6822? 

To try to answer these questions we follow two approaches: 1) We are
investigating the kinematical properties of C stars  in NGC 6822 from radial
velocity follow-up observations already acquired; 2)  We look at other 
dwarf galaxies similar or dissimilar to NGC 6822 to search
 for intermediate-age halos. In this regard, the
 natural comparison to NGC 6822 is 
NGC 3109, a galaxy of the same morphological type. 
In both galaxies, de Blok \& Walter (2000) and Barnes
\& de Blok (2001) have recently mapped extensive HI envelopes. 

NGC 3109, a Magellanic dwarf on the outskirts of the Local Group, is a
spectacular object seen through a large telescope. With a diameter of
nearly 30$'$, NGC 3109 is one of the largest and  brightest galaxy 
of the southern hemisphere. Well resolved into stars, it has been
classified Sm IV by Sandage \& Tammann (1981). A prime focus CTIO 4~m
photo of this galaxy can be found in Demers et al. (1985). 
This galaxy, first believed to be irregular because of the lack of an
obvious nucleus, is actually a disk galaxy and could possibly be 
a small spiral (Demers et al. 1985). 

During the last decade several photometric investigations of small 
areas in NGC 3109 have been published. We now have a fairly good
estimate of the distance of this galaxy: Musella et al. 
(1997) derive $(m-M)_0 = 25.67\pm 0.16$ from 24 Cepheids while
Minniti et al. (1999) obtain $(m-M)_0 = 25.62\pm 0.1$
from the I magnitude of the tip of the red giant branch (TRGB). More recently,
M\'endez et al. (2002) re-determined distances to out-lying members of
the Local Group using the TRGB method. For NGC 3109 they quote 
 $(m-M)_0 = 25.52\pm 0.06$. We adopt,
for our investigation, the weighted mean of these three estimates
$(m-M)_0 = 25.56 \pm 0.05$. The colour excess and the amount of
Galactic extinction toward
NGC 3109 are small. Schlegel et al. (1998) obtain
E(B--V) = 0.04, from the COBE and IRAS maps at 100 $\mu$m. This
corresponds to E(R--I) = 0.03 and A$_I$ = 0.054 using the
reddening ratios of Rieke \& Lebofsky (1985). This low reddening is
consistent with the colour of the ridge of bright stars, seen in
Figure 1.

The photometric study of a $9'\times 9'$ region, centered on NGC 3109,
by Minniti et al. (1999) revealed that this galaxy  possesses a 
Population II halo extending to $\sim$ 4.5$'$ (1.8 kpc) above and
below the plane of the galaxy. These authors determined, from the
position of the red giant branch on the CMD, that the metallicity of
that population is [Fe/H] = --1.8 $\pm$ 0.2. They also noted the
presence of red asymptotic giant branch stars (AGB), presumably C stars, in the main
body of the galaxy. This metallicity estimate roughly confirms the abundance
of --1.6 $\pm$ 0.2 determined in the same way by Lee (1993).
M\'endez et al. (2002) obtained [Fe/H] = --1.69 $\pm$ 0.06 also from the
colour of the red giant branch. We adopt, for our discussion, the
weighted mean value of [Fe/H] = --1.7.

\section{Observations}
The results presented here are based on observations obtained, in Service
Queue observing mode in March 2002, with the
CFH12K camera installed at the prime focus of the 3.66~m Canada-France-Hawaii
Telescope. The camera consists in a 12$k\times 8k$ pixel mosaic covering
a field of $42'\times 28'$, each pixel corresponding to 0.206 arcsec.
Images were obtained through Mould I and R filters and narrowband CN and
TiO filters, centered at 808.6 nm and 768.9 nm, respectively. 
A summary of the acquired data is presented in Table 1. The
total exposure time on NGC 3109 was 6 hours, under excellent seeing. 

   \begin{table}
      \caption[]{Summary of the observations}
    $$
       \begin{array}{llcc}
            \hline
            \noalign{\smallskip}
            Filter&exp.\ times&average\ seeing\ ('')& mean\ airmass  \\
           \noalign{\smallskip}
            \hline
            \noalign{\smallskip}
R&8\times 400\ s&0.79&1.534\\
I&6\times 360\ s&0.69&1.540\\
CN&7\times 1167\ s&0.70&1.571\\
TiO&7\times 1167\ s&0.76&1.629\\

            \noalign{\smallskip}
            \hline
         \end{array}
     $$
   \end{table}

The data distributed by the CFHT have been detrended. This means that the
images have already been corrected with the master darks, biases, and
flats. Fringes have been removed on I exposures under 60 seconds and
large scale structures such as the ``Skyring'' effect have been removed
when relevant. This pre-analysis produces 12 CCD images, of a given 
mosaic,  with the same zero point and magnitude scale. 

We carefully combine images taken with the
same filter, making sure that the final FWHM was similar to the average
FWHM given in Table 1. 
The photometric reductions were done by fitting model point-spread functions
(PSFs) using DAOPHOT-II/ALLSTAR series of programs (Stetson 1987, 1994).
Instrumental magnitudes are calibrated using equations provided by the
CFHT QSO team. The equations for R and I magnitudes are:

$$ R = 26.190 + m_r - 0.09(X_r - 1) + 0.0094(R-I),$$
$$ I = 26.185 + m_i - 0.04(X_i - 1 ) -0.0511(R-I),$$

where X$_r$ and X$_i$ are the airmasses of the R and I exposures.
For the instrumental magnitudes r and i, obtained with DAOPHOT, and for the
given exposure times, $m_r$ and $m_i$ correspond to:
$$ m_r = (r - 25.0) + 2.5\log (400),$$
$$ m_i = (i - 25.0) + 2.5\log (360). $$
By subtracting the above R and I equations and using the appropriate
airmasses we obtain:
$$ (R-I) = 0.099 + 1.064(r-i),$$
$$ I = i + 7.5542 - 0.0511(R-I).$$

   \begin{figure*}
   \centering
\includegraphics[width=10cm]{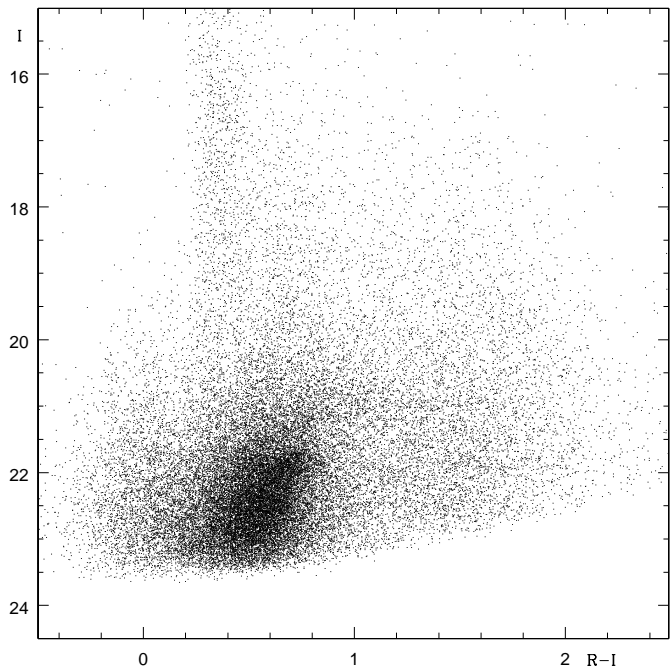}
   \caption{Colour-magnitude diagram of the whole CFH12K field. Stars with
colour errors $< 0.10$ are plotted. We acquire the first 1.5 magnitudes
of the giant branch, sufficient to identify the C star population.
               }
              \label{FigCMD}
    \end{figure*}

\section{Results}
The colour-magnitude diagram (CMD) of the whole CFH12K field is
presented in Figure 1. 38600 stars, with (R--I) colours with photometric
errors less than 0.1 mag. are plotted. Since our observations were
acquired to survey the C star population, they reach only
1.5 mag below the tip of the red giant branch. This limit is quite sufficient
to detect  all the C stars that can be resolved. 

The C stars are selected from the colour-colour diagram where we
plot their (CN--TiO) index versus their (R--I) color. Throughout  this
series of papers we have selected as an adoption criterion, stars
with $\sigma < 0.125$, where  $\sigma = (e_{R-I}^2 + e_{CN-TiO}^2)^{1/2}$.
In general this is not a magnitude cutoff criterion, but depends on
the quality of the (CN--TiO) index whose errors are  always the largest.
Experience has shown that these narrow filters
require exposures three to four times longer than I exposures. As shown 
in Table 1  the CN and TiO exposures are  3.78 times longer than I. 
Some 24000 stars satisfy this criterion. They
are plotted on the colour-colour diagram presented in Figure 2.
The zero point of the (CN--TiO) index is set according to the procedure
outlined by Brewer et al.  (1995). We set the mean of
(CN--TiO) = 0.0 for all blue stars since hot stars are expected to have
a featureless spectra in the CN and TiO regions. We thus define a blue star,
as in Letarte et al. (2002), i.e.  as a star
in the colour range $0.0 < (R - I)_0 < 0.45$. This of course requires an estimate
 of the
local colour excess which is adopted in the Introduction.
We define C and M stars as stars with (R--I)$_0$ $>$ 0.90: 
C stars have (CN--TiO) $>$ 0.3, while 
(CN--TiO) $<$ 0.0 for M stars. The C star
branch obviously extends to bluer colours. These C stars have fainter
magnitudes (Demers \& Battinelli 2002) than those inside the box. 
The limit (R--I)$_0$ = 0.90 corresponds to spectral type M0.
From this diagram we count, in the box,  446 C stars.

   \begin{figure*}
   \centering
\includegraphics[width=9cm]{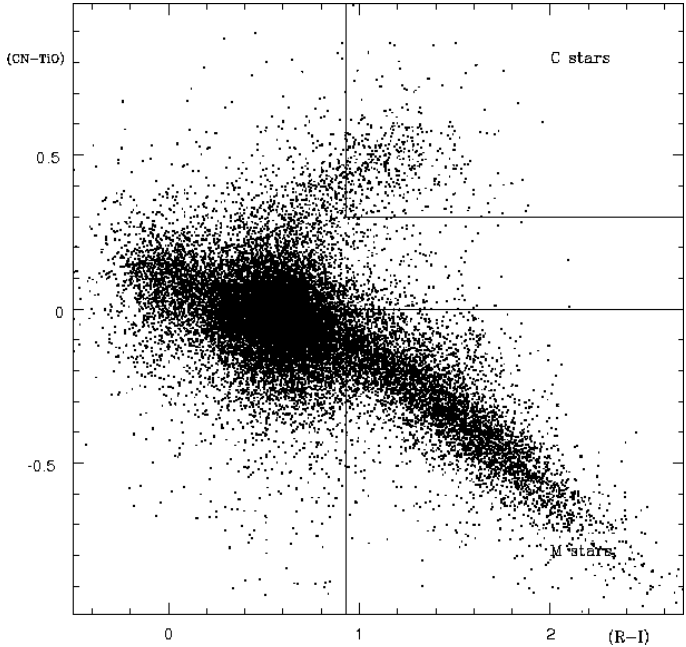}
   \caption{Colour-colour diagram showing the boxes of C stars and M stars.
               }
              \label{Figcc}
    \end{figure*}
Contrary to similar diagrams of other Local Group galaxies, already
published by us, the NGC 3109 colour-colour diagram shows substantially more scatter. 
We see many points where one does not expect to see any, for example, 
below the central concentration on the blue side of the M box.
This aspect of the diagram is explained by the poor quality of the 
photometry of the (CN -- TiO) index. Indeed, 57\% of the stars lying in 
the ``forbidden'' regions of the colour-colour diagram have (CN--TiO)
photometric errors larger than 0.075 while such fraction is  only 7\% among the identified
C stars.  We conclude from this scatter that the CN and TiO exposures should
have been somewhat longer. 

We list, in Table 2, the J2000.0 coordinates of the C stars identified
along with their magnitude and colours.
The 446 C stars are identified, by large dots, in the mosaic of the field
in Figure 3. From this figure, it is obvious that NGC 3109
does not possess an  extended intermediate-age halo, like the one seen around
NGC 6822 by Letarte et al. (2002). Thirty of the 446 C star
candidates (7\%) have photometric errors, for their (CN--TiO) index
larger than 0.075. They are represented, on Fig. 3, by the open squares.
We believe that these stars are probably not genuine C stars. They
could be eliminated from our list by tightening up our error criterion but
we prefer to keep the selection criterion at 0.125 to be coherent with other
papers of the series. 

  \begin{table}
      \caption[]{C stars in NGC 3109$^{\mathrm{a}}$}
    $$
       \begin{array}{lcccccccc}
            \hline
            \noalign{\smallskip}
            id&RA&Dec&I&\sigma_I&R-I&\sigma_{R-I}&CN-TiO&\sigma_{CN-TiO}  \\
           \noalign{\smallskip}
            \hline
            \noalign{\smallskip}
  1& 10:01:32.50& -26:15:10.50& 20.887&  0.032&  1.005&  0.053&  0.354&  0.076 \\
     2&10:01:34.32& -26:19:51.40& 21.189&  0.060&  1.829&  0.083&  0.669&  0.093 \\
     3&10:01:36.11& -26:21:02.20& 20.606&  0.025&  1.420&  0.037&  0.402&  0.055 \\
     4&10:01:45.01& -26:19:38.10& 21.206&  0.050&  1.101&  0.072&  0.317&  0.087 \\
     5&10:01:49.06& -26:22:54.60& 20.926&  0.025&  1.383&  0.057&  0.464&  0.096 \\
     6&10:01:52.62& -26:22:33.60& 22.559&  0.049&  1.022&  0.073&  0.322&  0.099 \\
     7&10:01:56.71& -26:03:07.70& 21.427&  0.060&  1.012&  0.081&  0.302&  0.076 \\
     8&10:01:57.68& -26:14:03.60& 19.564&  0.080&  1.392&  0.086&  0.417&  0.076 \\
     9&10:01:59.32& -26:03:52.30& 20.438&  0.031&  1.055&  0.039&  0.475&  0.033 \\
    10&10:02:07.77& -26:13:23.50& 19.947&  0.039&  1.311&  0.047&  0.639&  0.075 \\
    11&10:02:10.63& -26:07:17.60& 20.690&  0.009&  1.161&  0.013&  0.397&  0.019 \\
    12&10:02:11.44& -26:08:56.20& 20.922&  0.011&  1.501&  0.019&  0.375&  0.024 \\
            \noalign{\smallskip}
            \hline
         \end{array}
     $$
\begin{list}{}{}
\item[$^{\mathrm{a}}$] Complete Table 2 is 
available in electronic form at the CDS via anonymous ftp to cdsarc.u-strasbg.fr
(130.79.128.5) or via http://cdsweb.u-strasbg.fr/cgi-bin/qcat?J/A+A.
A portion is shown here for guidance regarding its
form and content. Units of right ascensions are hours, minutes and seconds, and
units of declination are degrees, arcminutes and arcseconds.

\end{list}
   \end{table}

   \begin{figure*}
   \centering
\includegraphics[angle=-90, width=13cm]{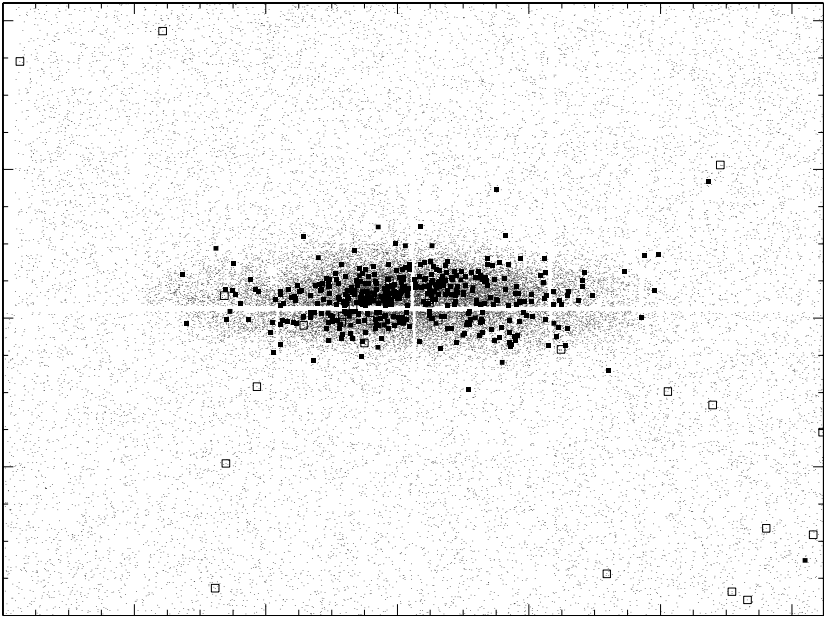}
   \caption{ The CFH12K field of NGC 3109. C stars identified
in this study are represented by filled squares. The open squares 
are C stars with photometric errors on their (CN--TiO) index $>$ 0.075.
North is on top, East is on the left.
The average gap, between the CCDs, in the X direction is 7.8$''$
while it is 6.8$''$ in the Y direction.
               }
              \label{Figmosaic}
    \end{figure*}
\section{Discussion}
\subsection{The photometric properties of C stars}
The I magnitude distribution of the C stars is displayed in Figure 4.
The shaded areas corresponds to the 30 stars with extreme (CN--TiO)
errors. The three faintest stars of our sample are among this subset but
the other ones are found at any magnitudes. Over a third of the stars
in this subset have (CN--TiO)$\approx$ 0.3, very near the border of the
C star box. 
 
The mean I magnitude for the 446 stars is $\langle I \rangle$ = 20.93,
while for the 416 C stars of better photometric quality 
$\langle I \rangle$ = 20.91, 
corresponding to a mean absolute magnitude of $\langle M_I \rangle$ = --4.71
a value, within the uncertainty of the distances,
 identical to the mean found for C stars in NGC 6822 (Letarte et al. 2002) 
and IC 1613 (Albert, et al. 2000). 
Their mean colour is $\langle (R-I)_0 \rangle$ = 1.17.

   \begin{figure*}
   \centering
\includegraphics[width=7cm]{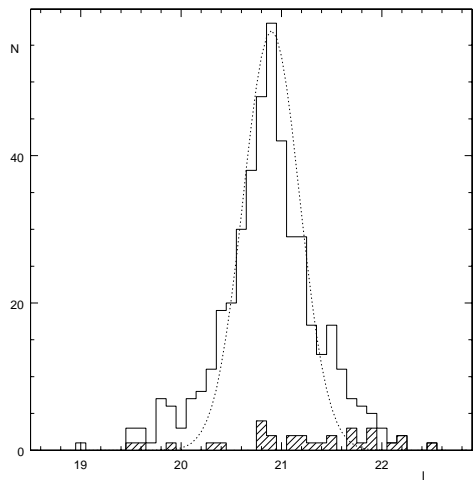}
   \caption{I magnitude distribution of the NGC 3109 C stars. The shaded
histogram corresponds to stars represented by open squares in Figure 3.
A  Gaussian with $\sigma = 0.28$ fits the central part of the distribution.
               }
              \label{Fighis}
    \end{figure*}
Figure 5 presents a comparison of  the $\langle M_I \rangle$ of C star 
population of various galaxies. The majority of the data points are
taken from our own investigations, with the exception of the two Magellanic
Clouds where we use the  DENIS data, for the SMC and data 
from Costa \& Frogel (1996) for the LMC. In both cases we select only
stars in our predefined colour range. For the DENIS data we adopt the
near infrared colour limits of 1.4 $<$ J--K $<$ 2.0 which is equivalent
to our R--I limits (Demers, et al. 2002). 
 The error bars reflect the uncertainties
in distance determinations. For Local Group galaxies this is usually
$\approx$ 0.1 mag. Since the publication of our results for SgrDIG,
(Demers \& Battinelli 2002), a new distance estimate by Momany et al. (2002)
brings the galaxy 0.1 mag closer. This is a galaxy, with 16 C stars, that
deviates most from the mean.
Galaxies with few ($<$ 40) C stars are identified
by open circles. The larger galaxies with hundreds of C stars are
represented by solid squares. Galaxies with numerous C stars show very
little dispersion in the $\langle M_I \rangle$. This suggests that
C stars could eventually be found to be useful standard candles.
If indeed the $\langle M_I \rangle$ of C stars is observationally
found to be independent of the metallicity of the parent galaxy, this
approach would be more interesting than the TRGB technique which requires
an estimate of the metallicity. Furthermore, 
contrary to Cepheids, C stars require observations at just one 
epoch. Our ongoing investigation should establish the range of galactic
properties where C stars could be used as distance indicators. Of
particular interest in this respect is the dichotomy between 
dwarf elliptical galaxies with no recent star formation  and dwarf
irregulars with recent star formation. Extreme cases could then be IC 10,
as an example of a starburst galaxy, 
and NGC 147 with NGC 185 as dwarf ellipticals big enough to contain 
several hundred C stars. The twin satellites of M31 have recently been
surveyed by Nowotny et al. (2003) who found  the $\langle M_i \rangle$
of the C star population to be substantially fainter than Fig. 5 would
suggest. Our recent investigation of NGC 147 (Battinelli \& Demers 2003)
does not support their findings.

   \begin{figure*}
   \centering
\includegraphics[width=11cm]{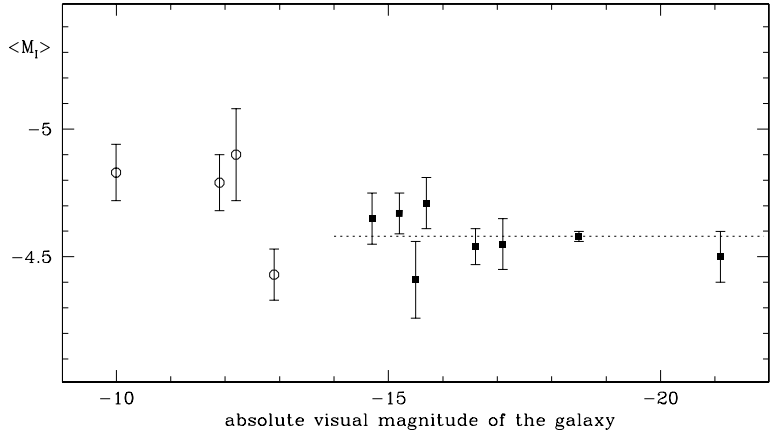}
   \caption{Comparison of the mean absolute magnitude of C stars in various
galaxies of the Local Group. Galaxies with few C stars are represented by
open circles while galaxies with  more than one hundred C stars are 
assigned solid dots. The mean $\langle M_I \rangle$ for bright galaxies
is --4.59.
               }
              \label{FigMI}
    \end{figure*}
\subsection{The structure of NGC 3109}

Our wide field mosaic of NGC 3109, reaching M$_I = -2$, can be used to
determine the scale length of the major or minor axis and to study
the distribution of the intermediate-age population. 
From the surface photometry of NGC 3109, Jobin \& Carignan (1990) established
several of its properties: The inclination of the disk from I photometry
is taken to be $i = 75$ $\pm 2^\circ$ and the position angle of the
disk to be $\theta_o = 93$ $\pm 2^\circ$. 
To facilitate  our analysis, we assume that the position angle of the
disk of NGC 3109 is exactly 90 degrees thus is aligned East - West.

\subsubsection{The major axis profile}
We determine the scale length of the disk by counting stars along the
major axis, using bins of 150$\times$1600 pixels, each having 2.719 arcmin$^2$.
The width in the north-south direction corresponds approximately to
the FWHM of the minor axis profile, described in the next section. 
A robust foreground surface density is evaluated by counting stars in 
two 1000 pixel
wide strips, located 500 pixels from the southern and northern borders of
the field. 
A total of 3860 stars, in these strips, yields a foreground density
of 13.75 $\pm$ 0.22 stars per arcmin$^2$.

The major axis profile is displayed in Figure 6. Counts on the eastern and
western sides are averaged and the foreground density is removed from those
counts. We presume that in the central
area the incompleteness is severe due to crowding. The straight line
corresponds to a linear regression from 4$'$ to 14$'$. The slope of
the line yields an exponential scale length of 2.96 $\pm$ $0.14'$, 
corresponding to 1.1 kpc, in excellent agreement with 
3.1$'$, determined by Jobin \& Carignan (1990). 

The discontinuity observed at 14$'$ must correspond to the end of the
disk and not be the result of an wrong estimation of the foreground density.
That limit corresponds to 4.7 times the disk scale length.
Kregel et al. (2002) determined, for 
small scale length spirals, that the average ratio of disk truncation
radius and scale length is four. Thus, NGC 3109 is quite normal in
this respect.


   \begin{figure}
   \centering
   \includegraphics[width=7cm]{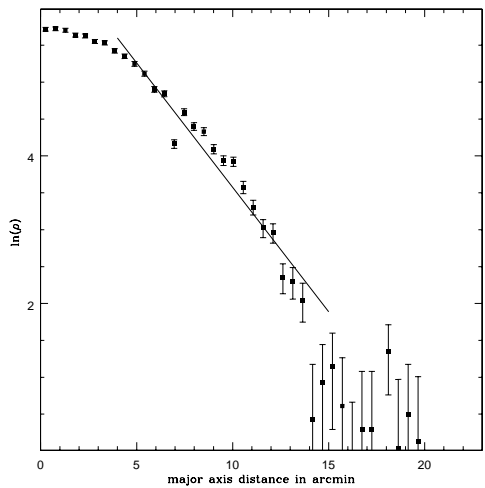}
      \caption{Density profile along the major axis of NGC 3109. We
average the eastern and western sides.}
         \label{Figmajprof}
   \end{figure}
%
\subsubsection{The minor axis profile}
In order to better define the minor axis profile and its limit, we count
only AGB and old red giants stars. To do so,  we exclude, from the
counts, stars with (R--I) $<$ 0.2 and obvious foreground stars with
I $<$ 19 mag. 
Stars are counted in a 2000 pixel N - S strip centered on
the galaxy. A foreground density of 11.88 $\pm$ $0.21$ stars per arcmin$^2$ is
determined by counting stars in the two strips described in the previous
section. This foreground
density differs from the one determined previously  because
here we consider a subset of the stars, as define above. The surface
density of C stars along the minor axis 
is also determined in the same fashion.

Figure 7 presents the surface density (stars/arcmin$^2$) along the 
minor axis, uncorrected
for the foreground,  at distances
larger than 2$'$ from the center of NGC 3109. This figure reveals that
the counts reach the foreground density at a distance between $4' -  5'$ from
the major axis. The dotted line corresponds to the foreground estimate,
made independently of the plotted data points. If one assumes that NGC 3109
is shaped like a thin disk with a 75$^\circ$ inclination, a 14$'$ semi-major
axis will correspond to a 3.75$'$ semi-minor axis, close to where we
observe the minor axis cutoff. The disk of the galaxy is however 
not infinitely thin thus  the minor axis cutoff is expected to be somewhat
larger than 3.75$'$.
This result shows that NGC 3109 does not have a halo made of red giants
of all ages, with M$_I < $--2.


   \begin{figure}
   \centering
   \includegraphics[width=7cm]{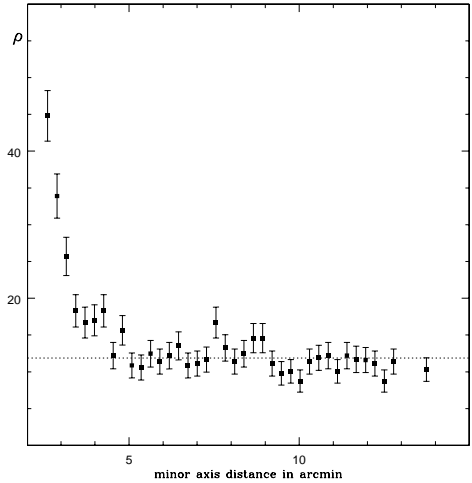}
      \caption{Surface density of red stars, star arcmin$^{-2}$  across the minor axis. 
The foreground density is reached between 4$'$ and 5$'$. }
         \label{Figminprof}
   \end{figure}
%

Two minor axis profiles are displayed in Figure 8: the red stars whose
profile is plotted in Fig. 7 and C stars represented by open dots. 
The C star counts
are multiplied by 20 to facilitate the comparison. A scale length of
0.576 $\pm$ $0.020'$ fits both the red star and C star
densities. There is a hint of  a short plateau or discontinuity
observed in the profile of the red stars in the last few bins. 
Counts at distances larger
than 4$'$ are very nearly equal to the foreground estimate. We interpret
this behavior as an indication of a real cutoff of the stellar population
of NGC 3109. The density of C stars is unaffected by the foreground
correction since it is zero for these stars 
but, unfortunately, their numbers are just too small to
allow us to follow them beyond $\sim 2.5'$. We have excluded, from this
analysis, the 30 C stars represented by open squares in Fig. 3.

   \begin{figure}
   \centering
   \includegraphics[width=7cm]{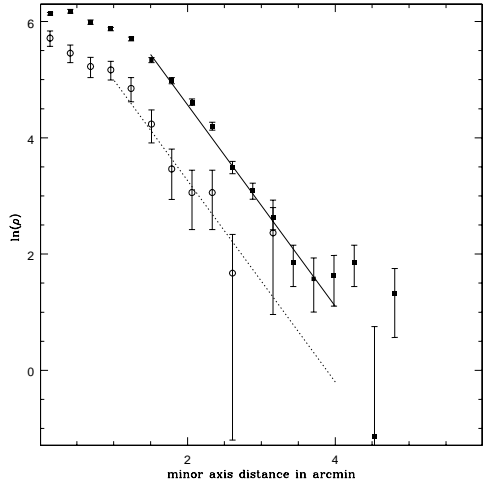}
      \caption{Density profiles across the minor axis. Solid dots
represent the AGB and red giant stars while the open symbols are
the C stars, their number multiplied by 20 to facilitate the comparison.}
         \label{Figminprof}
   \end{figure}
%

\subsection{The C/M ratio}
The C/M ratio of a stellar population is known to be function of the
metallicity of that population (Groenewegen 1999). The number of C stars
is relatively easy to obtain in a galaxy. On the other hand, 
the number of M stars is much more difficult to assess because one has to
know the contribution of the foreground M stars to the total count. 
The wide field of view of the CFH12K camera allows to obtain a robust
estimation of the foreground counts for galaxies of the size of NGC 3109.

To be consistent with our other publications of this series and to follow
Brewer et al. (1995), we select a subset of the M stars (those in the
M box of the colour-colour diagram) by applying a lower and upper cutoffs.
M$_{AGB}$ must have their M$_{bol} < -3.5$ and their I $>$ 19.0. The
determination of the bolometric magnitude follows the procedure detailed
by Battinelli et al. (2003). 

We define the ``galaxy'' by a rectangular area centered on the major axis
and with a width of 9$'$ and a length of 28$'$. In this area we count 413 C stars and 1318
M$_{AGB}$ stars. The foreground contribution is determined by counting
M$_{AGB}$ stars in two strips near the southern and northern borders of the
field. The foreground contribution to the M$_{AGB}$  counts of the rectangle
was found to be 1083 $\pm$ 37 stars. Therefore, 236 $\pm$ 51 M$_{AGB}$ stars
belong to NGC 3109, yielding a C/M ratio of 1.75 $\pm$ 0.20. 
The M star sample includes stars of spectral type M0 and later. 

\section{Conclusion}
Our  wide field survey of NGC 3109 has revealed 
marked differences with NGC 6822, an other dwarf galaxy of similar
morphological type. In term of the C star population, NGC 6822 with 
nearly 900 C stars has more intermediate-age stars for its luminosity
than any other dwarf surveyed. NGC 3109, on the other hand, fits into
the general trend of N$_C$ versus galactic M$_v$ defined by 
NGC 205, NGC 147 or IC 1613, galaxies of similar magnitude. Both
NGC 3109 and NGC 6822 are surrounded by an extended  hydrogen cloud.
This is essentially their only common structural feature. C star survey
of NGC 6822 has revealed the presence of an extended intermediate-age
halo as well as an old halo around that galaxy (Letarte et al. 2002).
NGC 3109, contrasts with the latter having no halo of intermediate-age
and no stellar halo outside of the (thick) disk. This properties makes
NGC 3109 unique among dwarf galaxies and justifies the title
of Jobin \& Carignan (1990) paper: The dark side of NGC 3109. 

\begin{acknowledgements}
This research
is funded in parts (S. D.) by the Natural Science and Engineering Council
of Canada.
\end{acknowledgements}

\end{document}